\documentclass[useAMS,usenatbib]{mn2e}
\usepackage{times,graphicx,amssymb,natbib}

\newcommand{\be}{\begin{equation}}
\newcommand{\e}{\end{equation}}
\newcommand{\bear}{\begin{eqnarray}}
\newcommand{\ear}{\end{eqnarray}}

\newcommand{\f}{\frac}
\newcommand{\de}{{\rm d}}

\begin{document}

\title[Escape fraction of photons from high redshift galaxies]
{The escape fraction of ionizing photons from high redshift galaxies from data-constrained reionization models}
\author[Mitra, Ferrara \& Choudhury]
{Sourav Mitra$^1$\thanks{E-mail: smitra@hri.res.in},~
Andrea Ferrara$^2$\thanks{E-mail: andrea.ferrara@sns.it}
and
T. Roy Choudhury$^3$\thanks{E-mail: tirth@ncra.tifr.res.in}~\\
$^1$Harish-Chandra Research Institute, Chhatnag Road, Jhusi, Allahabad 211019, India\\
$^2$Scuola Normale Superiore, Piazza dei Cavalieri 7, 56126 Pisa, Italy\\
$^3$National Centre for Radio Astrophysics, TIFR, Post Bag 3, Ganeshkhind, Pune 411007, India
} 

\maketitle

\date{\today}

\begin{abstract}
The escape fraction, $f_{\rm esc}$, of ionizing photons from high-redshift galaxies is a key parameter to
understand cosmic reionization and star formation history. Yet, in spite of many efforts, it remains largely 
uncertain. We propose a novel, semi-empirical approach based on a simultaneous match of the most recently
determined Luminosity Functions (LF) of galaxies in the redshift range $6\leq z \leq 10$ with reionization 
models constrained by a large variety of experimental data. From this procedure we obtain the evolution of the best-fit 
values of $f_{\rm esc}$ along with their 2-$\sigma$ limits. We find that, averaged over the galaxy population,  (i) the 
escape fraction increases from $f_{\rm esc} =0.068_{-0.047}^{+0.054}$ at $z=6$ to $f_{\rm esc} =0.179_{-0.132}^{+0.331}$ at 
$z=8$; (ii) at $z=10$ we can only put a lower limit  of $f_{\rm esc} > 0.146$. Thus, although errors are large, there is an 
indication of a 2.6 times increase of the average escape fraction from $z=6$ to $z=8$ which might partially release 
the ``starving reionization'' problem.
\end{abstract}


\section{Introduction}

One of the most crucial issues regarding the evolution of intergalactic medium (IGM) and cosmic reionization 
is the escape fraction, $f_{\rm esc}$, of ionizing photons from high-redshift galaxies.
This parameter remains poorly constrained in spite of the many theoretical and observational 
attempts made in past few years. The difficulties largely arise from the lack 
of a full understanding of the physics of star formation, radiative transfer and feedback processes, and from uncertainties on
the properties of the high-$z$ galaxy interstellar medium (ISM); as a result, derived values of $f_{\rm esc}$ span the large 
range $0.01-1$ \citep{2011ApJ...731...20F}.  Observationally, $f_{\rm esc}$ can be reliably estimated only at redshifts $z\lesssim3$ 
\citep{1995ApJ...454L..19L,2000ApJ...531..846D,2001ApJ...558...56H,2002MNRAS.331..463C,2002ApJ...568L...9G,2003MNRAS.342.1215F,2005A&A...435..471I,2006A&A...448..513B,2006ApJ...651..688S,2010ApJ...725.1011V}. 
On the other hand, theoretical studies \citep{2000ApJ...545...86W,2006ApJ...651L..89R,2008ApJ...673L...1G,2008ApJ...672..765G,2010PASA...27..110S,2010ApJ...710.1239R,2011MNRAS.412..411Y,2011arXiv1103.5226H,2011ApJ...731...20F,kuhlen2012} 
have been rather inconclusive so far, as illustrated by their often conflicting results in terms of $f_{\rm esc}$ values and trend with 
redshift and galaxy mass.  

One key aspect of reionization lies in its close coupling with the properties and evolution of first luminous sources 
(for reviews, see \citealt{2001ARA&A..39...19L,2001PhR...349..125B,2006astro.ph..3149C,2009arXiv0904.4596C}).
Observations of cosmic microwave background (CMB) and highest redshift QSOs put very tight constraints 
on the reionization history; these allow to construct self-consistent models of 
structure formation \citep{2005MNRAS.361..577C,2005ApJ...625....1W,2006MNRAS.370.1401G,2006MNRAS.371L..55C,2007MNRAS.379..253D,2007MNRAS.377..285S,2008MNRAS.391...63I,2011MNRAS.412.2781K}. 
The most favorable model, which is consistent with the Thomson scattering optical depth 
$\tau_{\rm el}=0.088\pm0.015$ from WMAP7 data \citep{arXiv:1001.4635v2} 
and the Gunn-Peterson optical depth evolution from QSO absorption line experiments at 
$z\gtrsim6$ \citep{2006AJ....132..117F}, suggests that reionization is an extended process 
over the redshift range $6\lesssim z\lesssim15$ \citep{2006MNRAS.371L..55C,mitra1,mitra2}. 
This model also indicates that reionization feeds back on star formation by suppressing it 
in the low-mass haloes at early times \citep{1996ApJ...465..608T,2007MNRAS.380L...6C}. 

In parallel, direct observations of galaxies at epochs close to the end of reionization have made astonishing progresses 
over the past few years \citep{2006NewAR..50..152B,2006Natur.443..186I,2007ApJ...670..928B,2008ApJ...686..230B,2008ApJ...677...12O,2009ApJ...705..936B,2009ApJ...697.1128H,2010ApJ...709L.133B,2010ApJ...709L..16O,2010arXiv1003.1706B,2010MNRAS.403..960M,2012ApJ...745..110O,2012arXiv1204.3641B}.  
allowing to derive the galaxy UV Luminosity Function (LF) up to $z\approx 10$ 
\citep{2006NewAR..50..152B,2010arXiv1006.4360B,2012ApJ...745..110O},
and to better constrain light production by reionization sources. 

Here we aim at combining data-constrained reionization histories and the evolution of the LF of early
galaxies to get an empirical determination of the escape fraction. The study also provides relatively tight 
constraints also on the evolution of the star-forming efficiency $\epsilon_*$\citep{fg2008,kuhlen2012}. 
Throughout the paper,  we assume a flat Universe with cosmological parameters  given by the 
WMAP7 best-fit values: $\Omega_m = 0.27$, $\Omega_{\Lambda} = 1 - \Omega_m$, $\Omega_b h^2 = 0.023$, and
$h=0.71$.  The  parameters defining the linear dark  matter power spectrum are $\sigma_8=0.81$, $ n_s=0.97$, 
$\de n_s/\de \ln k =0$ \citep{arXiv:1001.4635v2}. Unless mentioned, quoted errors are 2$\sigma$.

\section{Data-constrained reionization}
\label{sec:cfmodel}

We start by summarizing the main features of the semi-analytical model used in this work, which is based on 
\cite{2005MNRAS.361..577C} and  \cite{2006MNRAS.371L..55C}. 

The model follows the ionization and thermal histories of neutral, HII and HeIII 
regions simultaneously also accounting for IGM inhomogeneities described by a 
lognormal distribution as in \cite{2000ApJ...530....1M}. Sources of ionizing radiation are stars and quasars. 
The stellar sources, all characterized by a Salpeter IMF in the mass range $M_\star =1-100 M_{\odot}$,  can be divided into two classes, 
namely,  (i) metal-free (i.e. PopIII) stars;  (ii) PopII stars with sub-solar metallicities. The transition is 
based on a local critical metallicity criterion.  Radiative feedback, suppressing star formation in low-mass haloes, is included 
through a Jeans mass prescription based on the evolution of the thermal properties of the IGM. 

Given the collapsed fraction $f_{\rm coll}$ of dark matter haloes, the production rate of ionizing photons in the IGM is
\be
\dot{n}_{\rm ph}(z) = n_b
N_{\rm ion} \f{\de f_{\rm coll}}{\de t}
\e
where $n_b$ is the IGM number density, and $N_{\rm ion}$ is the number of photons entering the IGM per baryon included into stars. 
The parameter $N_{\rm ion}$ can actually be written as a combination of three parameters: the star-forming efficiency $\epsilon_*$ 
(fraction of baryons within collapsed halos going into stars), $f_{\rm esc}$, and the specific number of photons emitted per  
baryon in stars, $N_{\gamma}$, which depends on the stellar IMF and the corresponding stellar spectrum: 
\be
N_{\rm ion} = \epsilon_* f_{\rm esc} m_p \int_{\nu_{\rm HI}}^{\infty} \de \nu
\left[\f{\de N_{\nu}}{\de M_*}\right] = \epsilon_* f_{\rm esc} N_{\gamma}
\label{eq:nion}
\e
In our previous work (\citealt{mitra1,mitra2}), we assumed $N_{\rm ion}$ to be an unknown 
function of $z$ and decompose it into its  principal components. The Principal Component Analysis
filters out  components of the model that are most sensitive to the data and thus most accurately constrained.  
In the following we assume a single stellar population (PopII) when computing the ionizing radiation properties; 
any change in the characteristics of these stars  over time would be accounted for indirectly by the evolution of $N_{\rm ion}$. 
We also include the contribution of quasars at $z < 6$ assuming that they have negligible effects on IGM at 
higher redshifts; however, they are significant sources of ionizing photons at $z\lesssim 4$. 

\begin{figure}
\centering
\includegraphics[height=0.45\textwidth,width=0.4\textwidth, angle=270]{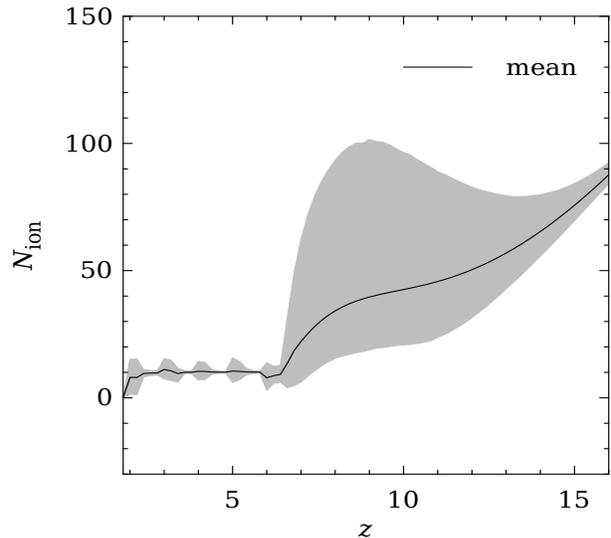}
\caption{Redshift evolution of $N_{\rm ion}$ obtained from the Principal Component Analysis using WMAP7 data. 
The solid line corresponds to the model described by mean values of the parameters while the 
shaded region corresponds to 2-$\sigma$ limits.}
\label{fig:NionPCA}
\end{figure}

From the above model, we obtain the redshift evolution of $N_{\rm ion}$ by doing a detailed likelihood analysis using three 
different data sets - the photoionization rates $\Gamma_{\rm PI}$ obtained using Ly$\alpha$ forest 
Gunn-Peterson optical depth observations and a large set of hydrodynamical 
simulations \citep{2007MNRAS.382..325B}, the redshift distribution of Lyman Limit Systems $\de N_{\rm LL}/\de z$ in 
$0.36 < z < 6$ \citep{2010ApJ...721.1448S} and the angular power spectra 
$C_l$ for TT, TE and EE modes using WMAP7 \citep{arXiv:1001.4635v2} and forecasted PLANCK data. 
We show the redshift evolution of $N_{\rm ion} (z)$ obtained from our Principal Component 
Analysis using WMAP7 data in Fig. \ref{fig:NionPCA}. The solid line corresponds to the model described by mean 
values of the parameters which we obtained by performing a Monte-Carlo Markov Chain (MCMC) 
analysis over the parameter space of our model, while the shaded region corresponds to its 2-$\sigma$ limits.
We concluded that it is not possible to match available reionization data with a constant 
$N_{\rm ion}$ over the whole redshift range (\citealt{mitra1,mitra2}). Rather, it must increase at $z > 6$ 
from its constant value at lower redshifts.This is a signature of either a varying IMF 
and/or evolution in the star-forming efficiency and/or photon escape fraction of galaxies, as eq. \ref{eq:nion} clearly shows.

At this point, it is worth pointing out some of the caveats in our formalism
based on PCA. The MCMC analysis for this work was done using 2--8 PCA
modes \citep{mitra2}.
Truncating a PCA expansion 
can reduce the variance in the estimation of the reionization history, but also introduces a bias towards the fiducial 
history. In order to account for this, we used the Akaike information criteria (AIC) to reduce the intrinsic bias present 
in any particular choice of fiducial model. We found that at $z\leq6$, the strong Lyman-$\alpha$ forest constraints 
essentially fix $N_{\rm ion}$, so that the efforts to reconstruct the reionization history are very promising at these redshifts. On the other hand, it is much more difficult to recover $N_{\rm ion}$ or the other various quantities related to 
reionization history at $z>6$ in a truly model-independent manner as there exists a considerable amount of bias at 
this high redshift end \citep{mitra1,mitra2}. It is possible that for these redshifts, the statistical uncertainties may have been hidden as systematic 
uncertainties \citep{2003PhRvL..90c1301H}.  However, with more data it would be possible to apply this 
technique in a regime where the variance in $N_{\rm ion}$ is small enough to produce a useful constraint on the 
reionization history without the need to truncate the PCA modes so severely and so without introducing any bias. This 
technique will become more applicable as more data becomes available for $z>6$ region.

\section{Luminosity function evolution}
\label{sec:lf_constraints}
The effect of reionization on the high redshift galaxy LF was studied using the 
semi-analytical models by \cite{2007MNRAS.377..285S} and \cite{2011MNRAS.412.2781K}. 
In this work, we follow their method to study the evolution of LF for our model. 

The LF is derived as follows. We compute the luminosity at 1500 $\mathring{\rm A}$ 
of a galaxy having the halo mass $M$ and age $\Delta t$ using 
\begin{equation}
 L_{1500} (M, \Delta t) = \epsilon_* \left(\frac{\Omega_b}{\Omega_m}\right)M l_{1500}(\Delta t)
 \label{eq:epsilon}
\end{equation}
Here the age of the galaxy formed at $z'$ and observed at $z$ is $\Delta t = t_{z}-t_{z'}$, 
$l_{1500}(\Delta t)$ is a template specific luminosity at 1500 $\mathring{\rm A}$ for 
the stellar population of age $\Delta t$. 
As we restrict to a single stellar population, i.e. PopII stars, $\epsilon_*$ 
indicates the star forming efficiency of PopII stars throughout the paper.

To compute $l_{1500}$, we use stellar population models of \cite{bc03} for PopII stars. 
The UV luminosity depends on galaxy properties including the IMF, 
star formation rate (SFR), stellar metallicity ($Z$) and age. 
\cite{dayal1} and \cite{dayal2} have shown that the metallicity correlates with 
stellar mass, and the best fit mass-metallicity relation they find is
\begin{equation}
 Z/Z_{\odot} = (0.25-0.05 \Delta z) \log_{10}(M_*)-(2.0-0.3 \Delta z)
 \label{eq:metallicity}
\end{equation}
where $\Delta z=(z-5.7)$ and $M_*$ is the total stellar mass of the galaxy. 
We take all the available stellar population models in the metallicity range $Z=0.0001-0.05$ 
for PopII stars and interpolate them to compute  $l_{1500}$ following the mass-metallicity relation 
given by the above relation for our model galaxies.

The luminosity can be converted to a standard absolute AB magnitude 
\citep{1983ApJ...266..713O,2007MNRAS.377..285S,2011MNRAS.412.2781K} using 
\begin{equation}
 M_{AB}=-2.5\log_{10}\left(\frac{L_{\nu_0}}{\rm erg \mbox{ } s^{-1} Hz^{-1}}\right)+51.60
\end{equation}
The luminosity function $\Phi (M_{AB},z)$ at any redshift $z$ is then given by 
\begin{equation}
 \Phi (M_{AB},z) = \frac{\de n}{\de M_{AB}} = \frac{\de n}{\de L_{1500}}\frac{\de L_{1500}}{\de M_{AB}},
\end{equation}
where 
\be
\f{\de n}{\de L_{1500}} = \int_{z}^{\infty}
\de z' \f{\de M}{\de L_{1500}}(L_{1500}, \Delta t)
\f{\de^2 n}{\de M \de z'}(M,z')
\e
is the comoving number of objects at redshift $z$ with observed luminosity within 
$[L_{1500}, L_{1500} + \de L_{1500}]$. The quantity $\de^2 n/\de M \de z'$ 
gives the formation rate of haloes  of mass $M$, which we obtain as in \cite{2007MNRAS.380L...6C}.
Note that, we can vary the star-forming efficiency $\epsilon_*$ in eq. \ref{eq:epsilon}, as 
a free parameter and obtain its best-fit value by comparing the high-redshift LFs computed 
using the above equations with observations. 

\subsection{Constraining the escape fraction} 

Our strategy to constrain $f_{\rm esc}$ exploits the combination between the previously derived (Sec. \ref{sec:cfmodel}) evolution
of $N_{\rm ion}$, and the constraints on $\epsilon_*$ that can be 
derived from matching LFs at different redshifts. Once the (Salpeter) IMF of the (PopII) stars is fixed, 
$N_\gamma$ is also fixed and equal to $\approx 3200$; from eq. \ref{eq:nion} we then get the value of  $f_{\rm esc}$ as follows:
\be
f_{\rm esc} = \frac{N_{\rm ion}}{\epsilon_* N_{\gamma}}
\label{eq:escfrac}
\e
As the uncertainties on $\left[N_{\rm ion}/N_{\gamma}\right]$ and $\epsilon_*$ are independent, 
the fractional uncertainty in $f_{\rm esc}$ can be obtained from the quadrature method \citep{taylor}), i.e. 
\begin{equation}
 \frac{\delta f_{\rm esc}}{f_{\rm esc}}=
\sqrt{\left(\frac{\delta\left[N_{\rm ion}/N_{\gamma}\right]}{\left[N_{\rm ion}/N_{\gamma}\right]}\right)^2+
\left(\frac{\delta\epsilon_*}{\epsilon_*}\right)^2}
\label{eq:erroresc}
\end{equation}
In this work, we are interested in the $z \ge 6$ evolution of the escape fraction.
In principle, our approach can also be used for the lower redshift range 
$3\leq z \leq 5$, provided that a detailed treatment of dust extinction 
is added to our model. The underlying assumption in the present work is that 
dust effects on the escape fraction can be safely neglected at early times. 

\section{Results}
\label{sec:results}

\begin{figure}
\centering
\includegraphics[height=0.26\textwidth,width=0.24\textwidth, angle=270]{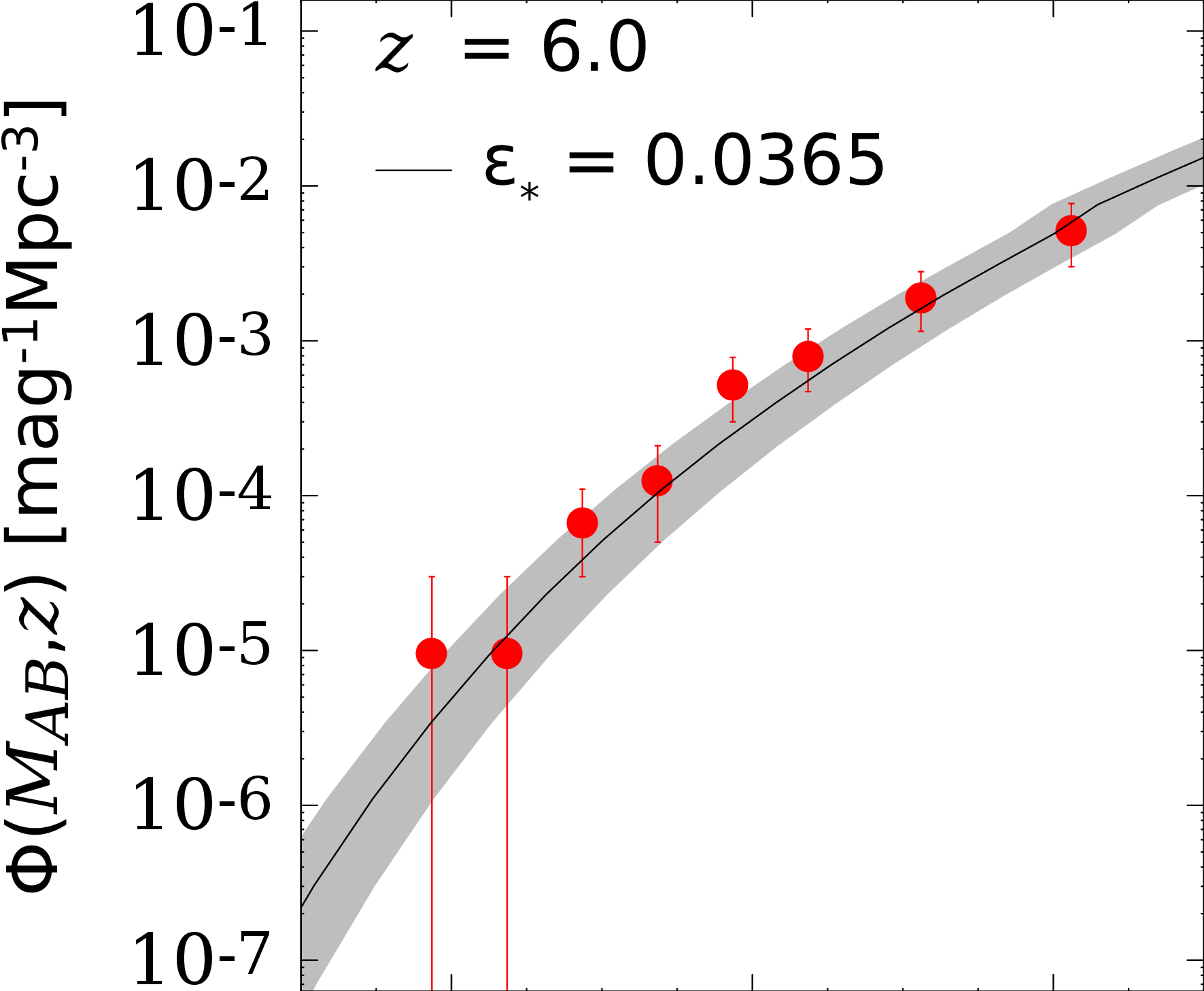}
\includegraphics[height=0.20\textwidth,width=0.24\textwidth, angle=270]{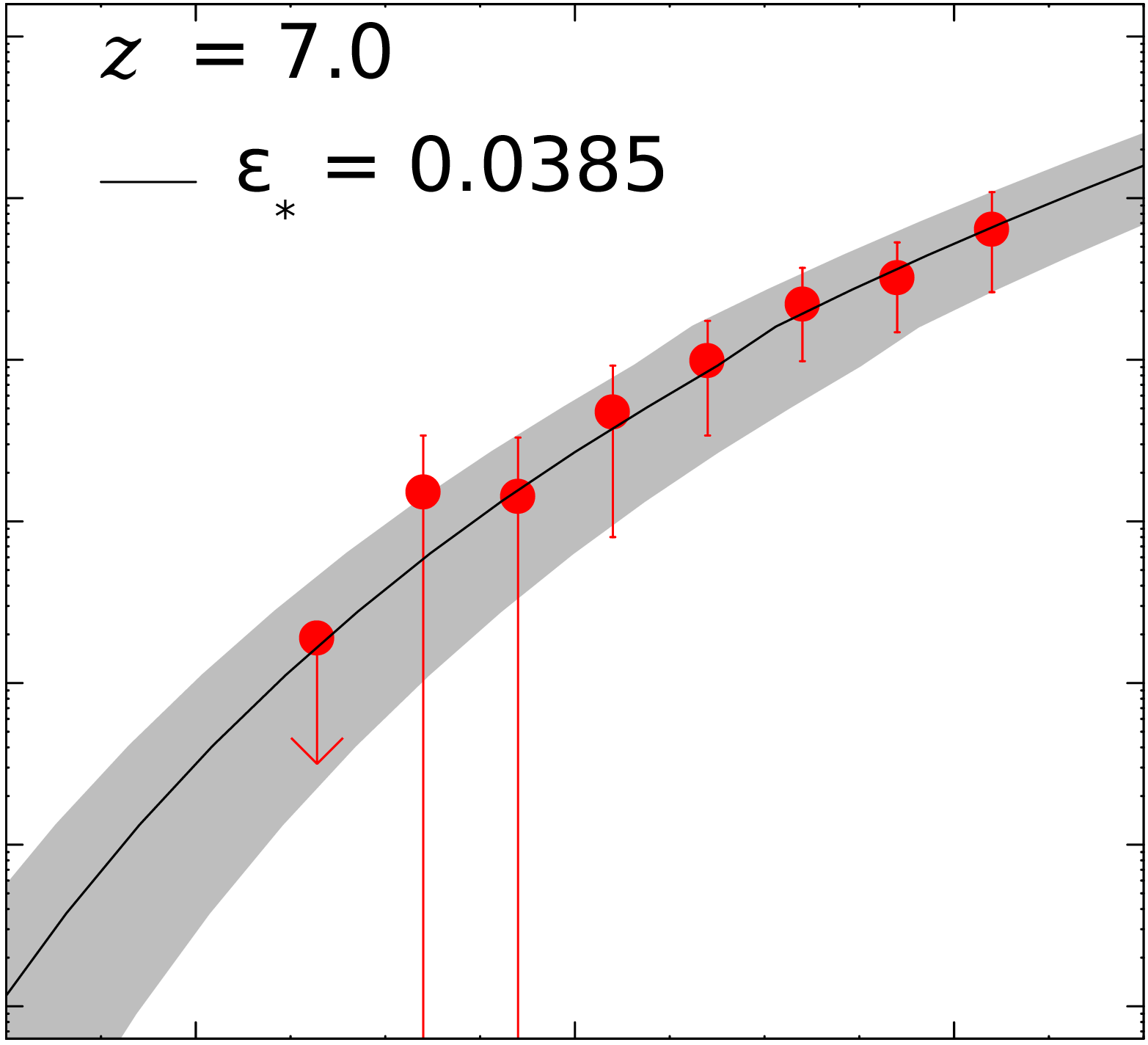} \\
\includegraphics[height=0.26\textwidth,width=0.289\textwidth, angle=270]{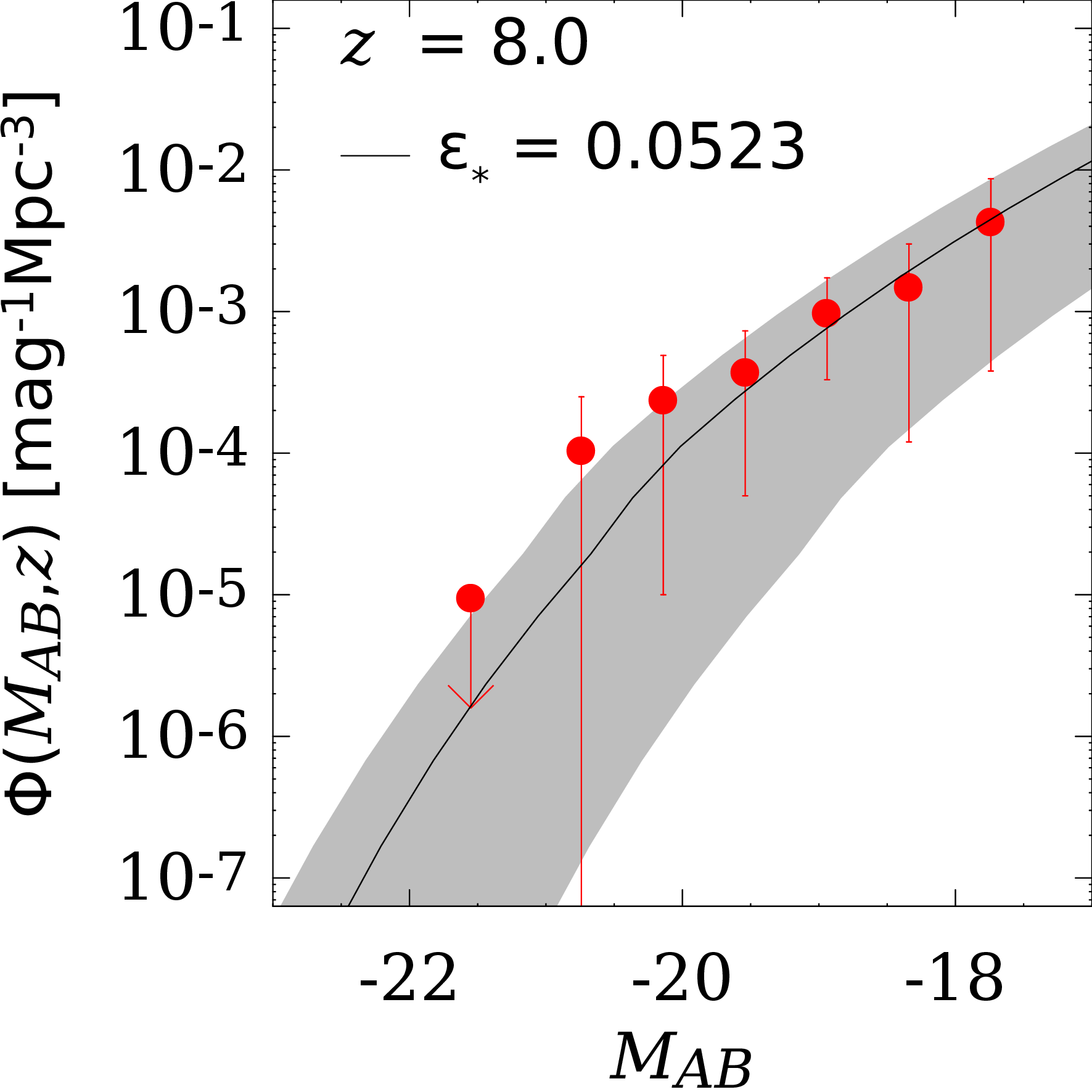}
\includegraphics[height=0.20\textwidth,width=0.289\textwidth, angle=270]{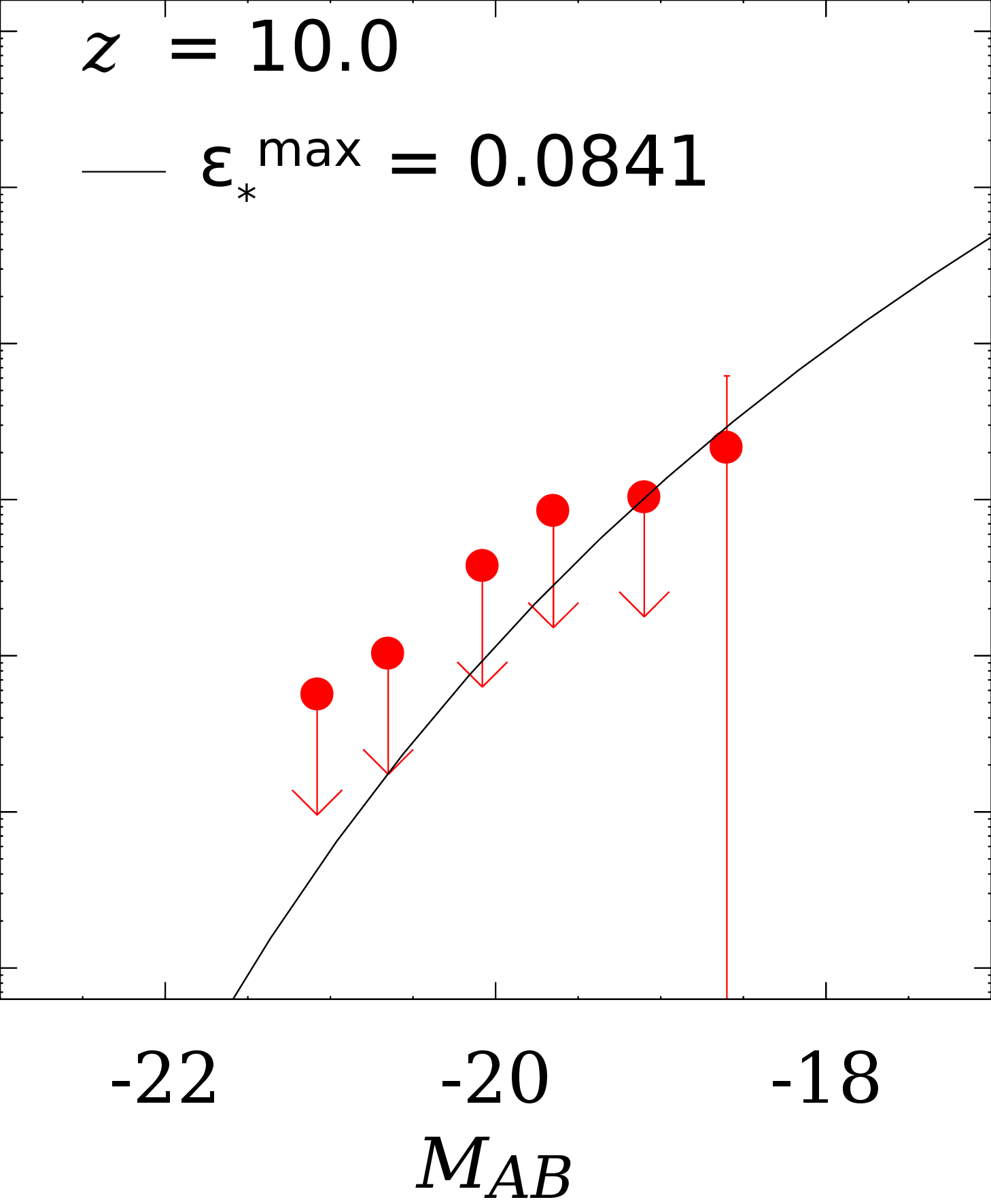}
\caption{Luminosity function from our model for best-fit $\epsilon_*$ (black curve) 
and its 2-$\sigma$ limits (shaded region) at $z=6$, 
$7$, $8$ and $10$. Data points with 2-$\sigma$ errors are from Bouwens \& 
Illingworth (2006) ($z=6$), Bouwens et al. (2010) ($z=7,8$) and Oesch et al. (2012) 
($z=10$). For $z=10$, we show the luminosity function from our model for the maximum 
value of $\epsilon_*$ for which the LF curve does not exceed the experimental upper limits.}
\label{fig:lumfunc}
\end{figure}

\begin{table*}
\begin{tabular}{c|c|c|c|c}
Redshift & Best-fit $\epsilon_*$ & 2-$\sigma$ limits & Best-fit $f_{\rm esc}$ & 2-$\sigma$ limits\\
\hline
$z=6$ & $0.0365$ & $[0.0253,0.0481]$ & $0.0684$ & $[0.0210,0.1221]$\\
$z=7$ & $0.0385$ & $[0.0193,0.0576]$ & $0.1607$ & $[0.0319,0.4451]$\\
$z=8$ & $0.0523$ & $[0.0129,0.0822]$ & $0.1794$ & $[0.0466,0.5098]$\\
$z=10$& $< 0.0841$ & &$> 0.1456$ & \\
\hline
\end{tabular}
\caption{Best-fit values and 2-$\sigma$ limits of $\epsilon_*$ and the derived parameter $f_{\rm esc}$ 
  for the reionization model obtained from the LF calculation at different redshifts. At $z=10$, 
  we get only an upper limit of $\epsilon_*$ and a corresponding lower limit of $f_{\rm esc}$.} 
\label{tab:escfrac}
\end{table*}

The observationally determined LFs are taken from \cite{2006NewAR..50..152B} 
for $z=6$, \cite{2010arXiv1006.4360B} for $z=7,8$ and \cite{2012ApJ...745..110O} 
for $z=10$. Figure \ref{fig:lumfunc}  shows the globally averaged LFs calculated using our model for 
$z=6,7,8,10$ compared to the observational data points. The $z=10$ data are 
obtained from the detection of a single galaxy candidate by \cite{2012ApJ...745..110O}; hence, 
we only show results for the maximum value of $\epsilon_*$ for which the LF curve does not exceed the 
experimental upper limits.
 
Our model reproduces  the observed LFs reasonably well, especially at lower redshifts. From 
such a match we find that the best-fit value of the star-formation efficiency $\epsilon_*$ 
nominally increases from 3.6\% at $z=6$ to 5.2\% at $z=8$. Such a small variation is 
statistically consistent with a constant value of $\epsilon_*$, i.e. no evolution.

\begin{figure}
\centering
\includegraphics[height=0.45\textwidth,width=0.4\textwidth, angle=270]{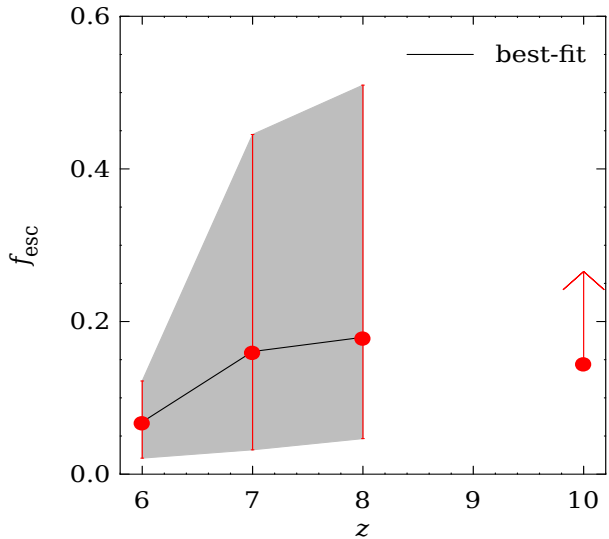}
\caption{Redshift evolution of the escape fraction $f_{\rm esc}$ with 2-$\sigma$ errors. The $z=10$ 
  data point shows the lower limit of $f_{\rm esc}$. The solid line corresponds to its best-fit value 
  while the shaded region corresponds to 2-$\sigma$ limits.} 
\label{fig:escfrac}
\end{figure}

The corresponding values of $f_{\rm esc}$ calculated using eq. \ref{eq:escfrac} and \ref{eq:erroresc} are
plotted in Fig. \ref{fig:escfrac} along with the 2-$\sigma$ confidence limits (shaded region). 
The numerical values for $\epsilon_*$ and $f_{\rm esc}$ are also reported in  Table \ref{tab:escfrac} 
for different redshifts ($z=6,7,8$). 
The escape fraction shows a moderately increasing trend from $f_{\rm esc} =0.068_{-0.047}^{+0.054}$ at 
$z=6$ to $f_{\rm esc} =0.179_{-0.132}^{+0.331}$ at $z=8$; at $z=10$ we can only put a lower limit  of 
$f_{\rm esc} > 0.146$, corresponding to the maximum allowed value of $\epsilon_*=0.0841$. 

%
The reported 2-$\sigma$ errors are however relatively large and we cannot exclude a
non-evolving galaxy-averaged trend for  $f_{\rm esc}$ . The uncertainties become larger
with redshift as a consequence of the fact that the larger LF errors at higher redshifts.

\section{Conclusions}
\label{sec:conclusions}

We have used a semi-analytical model, based on \cite{2005MNRAS.361..577C} and 
\cite{2006MNRAS.371L..55C} to compare the UV luminosity functions at different epochs 
predicted from our model with the observed LF to constrain the parameters 
related to star formation history in the redshift range $6\leq z \leq 10$. 
In particular, by varying the star formation efficiency as a free parameter, we have constrained 
one of the most unknown parameters of reionization models, the escape fraction 
$f_{\rm esc}$ of ionizing photons from high-redshift galaxies. The main findings of our work
are that, averaged over the galaxy population,  (i) the escape fraction shows a moderate increase from 
$f_{\rm esc} =0.068_{-0.047}^{+0.054}$ at $z=6$ to $f_{\rm esc} =0.179_{-0.132}^{+0.331}$ at 
$z=8$; (ii) at $z=10$ we can only put a lower limit  of $f_{\rm esc} > 0.146$. Thus, although errors are large, 
there is an indication of a 2.6 times increase of the average escape fraction from $z=6$ to $z=8$ which might 
partially release the ``starving reionization'' problem. At the same time, the best-fit value of the star formation 
efficiency $\epsilon_*$ nominally increases from 3.6\% at $z=6$ to 5.2\% at $z=8$. Such a small variation is 
statistically consistent with a constant value of $\epsilon_*$, i.e. no evolution.

Parallel to our more phenomenological approach, in the past few years many numerical and analytical studies 
have attempted to constrain $f_{\rm esc}$ reaching often contradictory conclusions, likely due to 
uncertainties on star formation history, feedback, radiation transfer and the geometry of the ISM distribution
\citep{2011ApJ...731...20F}. Increasing \citep{2006ApJ...651L..89R,2010ApJ...710.1239R,2011arXiv1103.5226H}, 
decreasing \citep{2000ApJ...545...86W} or un-evolving \citep{2008ApJ...673L...1G,2011MNRAS.412..411Y} trends have
been suggested as a function of redshift.

A strong redshift evolution of the escape fraction was recently found by \cite{kuhlen2012}. They show that, 
models in which star formation is strongly suppressed in low-mass haloes, can simultaneously satisfy reionization 
and lower redshift Lyman-$\alpha$ forest constraints only if the escape fraction of ionizing radiation 
increases from $\sim 4\%$ at $z=4$ to $\sim 1$ at higher redshifts. 
Although broadly in agreement with their conclusions, our results show instead that reionization and LF data can 
be satisfied simultaneously if  $f_{\rm esc}$ grows from $\sim 7\%$ at $z=6$ to $\sim 18\%$ at $z=8$, but without requiring 
an escape fraction of order of unity at these redshifts. We believe that this discrepancy can be understood as due to 
the fact that unlike \cite{kuhlen2012}, we are fitting the \emph{full CMB spectrum} rather than the single value of
$\tau_{\rm el}$; the latter choice can be thought as a simplification of CMB polarization observations. In addition, we have
used a PCA analysis to optimize model parameters to reionization data, yielding a more robust statistical analysis.

Although here we have only considered the evolution of $z\geq6$ luminosity functions, our approach can also be applied 
to model the LFs at $3\leq z \leq 5$. As hydrogen reionization mostly occurs at $z\gtrsim 6$, the LFs in this lower redshift 
range are very unlike to be sensitive to the details of reionization history. Also, dust extinction at $z<6$ can 
decrease $f_{\rm esc}$ by absorbing the ionizing photons at these epochs \citep{2011MNRAS.412..411Y}. 
As a caveat we mention that the present results can be responsive to changes in some cosmological parameters, mainly 
$\sigma_8$ and $n_s$ \citep{2011PhRvD..84l3522P}. A larger  $\sigma_8$ or $n_s$ may lead to an increase in the number of collapsed haloes at all 
redshifts. In principle then, one should include these two quantities in the analysis as additional free parameters. 
Also, it could be interesting  to evaluate the effects of PopIII stars and other feedback processes in our LF calculation. 
We hope to revisit some of these topics in more detail in future work.

\bibliographystyle{mn2e}
\bibliography{medea}

\end{document}